\documentstyle[aps,preprint]{revtex}
\begin{document}
\draft

\title{On the time evolution of the entropic index}
\author{Jin Yang $^{1}$, Paolo Grigolini$^{1,2,3}$, }
\address{$^{1}$Center for Nonlinear Science, University of North Texas,\\
P.O. Box 5368, Denton, Texas 76203 }
\address{$^{2}$Istituto di Biofisica del Consiglio Nazionale delle\\
Ricerche, Via San Lorenzo 26, 56127 Pisa, Italy }
\address{$^{3}$Dipartimento di Fisica dell'Universit\`{a} di Pisa, \\
Piazza\\
Torricelli 2, 56127 Pisa, Italy }
\date{\today}
\maketitle

\begin{abstract}
We adapt the Kolmogorov-Sinai entropy to the
 non-extensive perspective recently advocated by Tsallis. The
 resulting expression is an average on
 the invariant distribution, which should be used to
detect the genuine entropic index $Q$.  We argue
that the condition $Q >1$ is  time dependent.

\end{abstract}


The importance of establishing a connection between dynamics and
thermodynamics, and the difficulties with this ambitious purpose, are
illustrated in a very attractive way in the recent book
by Zaslavsky\cite{zaslavsky}. Zaslavsky shows that the problems with
this connection are not caused by the phenomenon of the
Poincar\'{e} recurrences:In strongly chaotic systems the Poincar\'{e}
recurrences are frequent
and erratic, and result
 in a Poisson form for the distribution of recurrence times
 $P_{R}(t)$, namely a distribution with the following form ($t>0$)
\begin{equation}
P_{R}(t) \propto \exp(-h_{KS}t)\ ,  \label{bernouilli}
\end{equation}
where $h_{KS}$ denotes the Kolmogorov-Sinai entropy\cite{KS}, thereby proving
the intimate connection between thermodynamics and mechanics in this case.
The problem is that Eq.(\ref{bernouilli}) rests on a condition
of strong chaos that seems to be an exception rather than a rule.
In general, the
phase space of Hamiltonian systems is rarely totally chaotic. We cannot rule
out the possibility that even in the case of seemingly chaotic systems
 small islands of
stability lie on the phase space. The presence of an island of stability
has  impressive consequences. The region of separation between the
deterministic island and the chaotic sea is fractal and self-similar. These
properties result in stikness.This means that a generic trajectory with
initial conditions located somewhere else, in the chaotic sea, through
a fast process of diffusion will reach that surface and will
stick to it for very extended sojourn times, with the distribution
density $\psi(t)$. As a consequence , we have:
\begin{equation}
\lim_{t\rightarrow \infty} P_{R}(t) =  \psi(t)\
\label{waitingdominance}
\end{equation}
and
\begin{equation}
\lim_{t\rightarrow \infty} \psi(t)  = const/t^{(2 + \beta)}\
\label{inversepower},
\end{equation}
with$\beta>1$. This results is compatible with the important theorem of
Kac \cite{KAC}, according to which the mean value of the first moment
of the distribution $P_{R}(t)$ is finite.

Zaslavsky\cite{zaslavsky} also points out that two distinct
billiards, each of them characterized by the dynamical properties
necessary to realize the ergodic condition,
when coupled to one another through a small hole in
the wall that separates one billiard from the other, do not show any
equilibration, at least for a very extended interval of time, and
rather the trajectories seem to prefer one of the
two billiards: An effect reminiscent of the action of Maxwell's demon. This
is the
consequence of the breakdown of the thermodynamic
condition of Eq.(\ref{bernouilli}), provoked by the emergence
of the slow tail of Eq.(\ref{inversepower}).
Zaslavsky\cite{zaslavsky} reaches the important conclusion ``that
chaotic dynamics exhibit some memory-type features which have to be
suppressed in order to derive the laws of thermodynamics''. In this
paper we want to prove that if such a memory erasing process exists,
it must be perceived as the source of  a transition from non-extensive to
extensive thermodynamics rather than a transition from dynamics to
thermodynamics.

First of all, let us stress that it is essential to use
the
Tsallis entropy \cite{CONSTANTINO88}
rather than the conventional Gibbs entropy.  The Tsallis
 entropy reads
 \begin{equation}
 H_{q}= \frac{1 - \int dx\Pi(x)^{q}}{q-1}.
 \label{entropy}
 \end{equation}
 Note that this entropy is characterized by the index $q$ whose
 departure from the
 conventional value $q = 1$ signals the thermodynamic effects of
 either long-range correlations in fractal dynamics\cite{brazil} or the
non-local
 character of quantum mechanics \cite{brazil,luigi}. The function
 $\Pi(x)$ denotes a distribution of a generic variable $x>0$,
 including the case where $x \equiv t$, with $t$ being a Poincar\'{e}
 recursion time or a time of sojourn
 at the border
 between chaotic sea and stability island. As earlier pointed
 out, this latter physical interpretation applies to the case where
 fractal dynamics result in a breakdown of the conventional
 condition of a finite time scale. It  has been recently
 pointed out\cite{ANNA} that the maximization of the entropy of
 Eq.(\ref{entropy}) under the condition of the existence of a finite
 first moment, in line with the  theorem of Kac\cite{KAC}, results in
 an inverse power law like that of Eq.(\ref{inversepower}).

 As important as this result is, it would leave open the problem raised by
 Zaslavsky as to the connection between dynamics and thermodynamics,
 if, as it is correct to do, a special attention were to be devoted to
 the Kolmogorov-Sinai (KS) entropy.
 The study of the connection between dynamics and thermodynamics is making
significant progresses along the lines of the seminal work of
Krylov~\cite{Krylov}. Under his influence interesting attempts are
currently being made at relating the
KS entropy to the thermodynamical entropy. Of
remarkable interest are the work of Gaspard, relating the KS entropy for a
dilute gas to the standard thermodynamical entropy per unit volume of an
ideal gas~\cite{Gaspard}, and the more recent paper by Dzugutov, Aurell and
Vulpiani~\cite{DAV98}, who express the KS entropy of a simple liquid in
terms of the excess entropy, namely, the difference between the
thermodynamical entropy and that of the ideal gas at the same thermodynamical
state. Latora and Baranger\cite{latora} studied the KS entropy of
some maps, and for one of them, the cat map, they found analytical
results. To properly appreciate the significance of the result
obtained by these authors, we must note first of all that the
KS entropy is a kind of entropy per unit of time, and
that within this perspective the thermodynamical regime is expected to
be expressed by a condition where the entropy growth
is linear in time. Latora and Baranger\cite{latora} found that at short times
the regime of entropy increase, instead of being
linear,  is exponential in time. They also found that
the KS regime, beginning at the end of this initial transient process,
 is not permanent and that after a given time a form of
saturation takes place. All these are indications of a time evolution
of the thermodynamical properties of the systems that are probably
related to the main problem under discussion in this paper, namely
the aging of the non-extensive thermodynamics of Tsallis \cite{CONSTANTINO88}.

A recent work by Tsallis, Plastino and Zheng\cite{TPZ97} illustrates
the convenience of generalizing the KS procedure so as to make it
efficient to study the thermodynamical properties of fractal dynamics.
However, these authors adopt heuristic arguments and do not provide direct
prescriptions on how to express the
ensuing generalized version of KS entropy in terms of the invariant
distribution. We refer to this generalized form of KS entropy as
Kolmogorov-Sinai-Tsallis (KST) entropy. Its explicitly form
reads:

\begin{equation}
H_{q}({N}) = \frac{1 -  \sum_{\omega_{0}\ldots\omega_{N - 1}}
p(\omega_{0}\ldots\omega_{N-1})^{q}}{q-1}.
\label{KST}
\end{equation}
The numerical procedure to
evaluate the KST entropy is the same as that we have to adopt to evaluate
the KS entropy. For clarity we remind the reader about this prescription
using the most elementary phase-space as possible, namely, a
one-dimensional interval $[0,1]$ for the continuous variable $x$.
This phase space is divided in $l$ cells with the same width $1/l$.
Then we run the dynamical system under study. In this paper we shall
focus on one-dimensional maps, thereby fitting the restriction of
considering a one-dimensional phase space. However,
our conclusions
are not restricted to maps, since, as we shall see, these conclusions
 can be straightforwardly
applied to the hamiltonian systems discussed by Zaslavsky\cite{zaslavsky}.

Running a one-dimensional map means producing a sequel of values
$x_{0}\ldots x_{j} \ldots $. Since any point of this one-dimensional trajectory
is located in a given cell, producing a trajectory is equivalent to
generating a sequel of values $\omega_{0}\ldots\omega_{j}\ldots$, where
$ \omega_{j}$ is the label of the cell occupied by the trajectory at the
``time'' $j$. After creating the sequence
$\omega_{0}\ldots\omega_{j}\ldots$, which for simplicity we assume to be
infinitely
long, we proceed with the evaluation of the KST entropy as follows.
We fix a window of size $N$ and we move the window along the sequence.
For any window position we record the labels lying within the window,
starting from the first, $ \omega_{0}$, till to last, $\omega_{N-1}$. Notice
that the subscrits now refer to the order within the window, which must
not be confused with the subscript denoting the position of the symbol
in the whole sequence. Moving the window we can evaluate how many times
the same combinations of symbols appears, and from this frequency we
evaluate the probablity distribution $p(\omega_{0}\ldots\omega_{N-1})$
which is then used to evaluate the KST entropy of
Eq.(\ref{KST}).

This kind of calculation has been recently \cite{buiatti} applied
to evaluating the KST
of a symbolic sequence obtained with a stochastic rule essentially
equivalent to the intermittent dynamical process behind the
one-dimensional processes of L\'{e}vy diffusion\cite{ALLEGRO}, and
consequently, equivalent, in principle, to the processes studied in
Ref.\cite{ANNA}. We notice that the result of the research work of
Ref.\cite{buiatti} proved that the KST entropy is a linear function of N
only when
the proper entropic index $q$ is used. This entropic index turned out to
fulfill the condition $q\leq 1$ in a striking conflict with the result of
Ref.\cite{ANNA}, which predicts $q > 1 $. This paper, among other things,
aims at shedding light into the reasons for this conflict, which will be duly
accounted for.

The key way of settling the problems of determining the KST adopted in
this paper is the same as that found by Pawelzik and Shuster\cite{PS87} to
evaluate a different form of generalized KS entropy, and precisely
that obtained by replacing the Gibbs entropy with the R\'{e}nyi
entropy\cite{renyi}, namely, the  Kolmogorov-Sinai-R\'{e}nyi (KSR)
entropy. For the calculation of both the  KST and KSR entropy it is
essential to adopt the following very important relation:
 \begin{equation}
 \sum_{i} p_{i}^{q}= \frac{1}{M}\sum_{j=1}^{M}\tilde{p}_{j}^{q-1}.
 \label{key}
 \end{equation}
 which has been advocated by Pawelzik and Schuster \cite{PS87}.

 This important relation is worth of some illustration. For any
 window of size $N$ we create a multidimensional phase space, of
 dimension $N$, associating any point $x_{j}$ of our original
 one-dimensional phase space with the $N$-dimensional point
 $x_{j},x_{j+1}\ldots x_{j+N-1}$. This means that the original
 one-dimensional cells of size $1/l$ become $N$-dimensional squares
 with the same size. Any cell of the resulting $N$-dimensional phase
 space corresponds to one of the combinations
 $\omega_{j},\omega_{j+1}\ldots\omega_{i+N-1}$ whose frequency must
 be properly evaluated to determine the distribution
 $p(\omega_{0}\ldots\omega_{N-1})$, which, in turn, is necessary
 for the calculation of the KST entropy of Eq.(\ref{KST}).The
cells of this N-dimensional phase-space are properly labelled and the
symbol $p_{i}$ appearing on the l.h.s. of Eq.(\ref{key}) is the
probability that a trajectory of this $N$-dimensional phase space
running for an unlimited amount of time is found in the $i-th$ cell. The
term on the r.h.s. of Eq.(\ref{key}) refers to a calculation procedure
based on the adoption of a single trajectory running from the time
$j=1$ to the time $j=M$. This trajectory carries with itself a
$N$-dimensional square of size $2/l$, of which the
 $N$-multidimensional point $x_{j},x_{j+1}\ldots x_{j+N-1}$
 is the center. The symbol $\tilde{p}_{j}$ denotes the probability
 that the same trajectory, at earlier or later times, is found in this
 cell. The mathematical arguments invoked by
 Pawelzik and Schuster\cite{PS87} to explain why the power $q-1$ on the r.h.s.
 of Eq.(\ref{key}) corresponds to the power $q$ appearing on the
 l.h.s of the same equation has a transparent physical meaning: The
 trajectory  carrying the moving $N$-dimensional cell of size $2/l$
 explores with more (less) frequency the regions of the $N$-dimensional phase
 space of higher (lower) probability, so that the sum over the running
 index {j} implicitly includes the missing
 factor of $\tilde{p}_{j}$.

 Rather than using the key relation of Eq.(\ref{key}) to introduce
 a generalized correlation integral, as Pawelzik and Schuster propose,
 we now have recourse to an approach inspired to that of
 Tsallis et al.\cite{TPZ97}. To see how this procedure works let us
 consider, as an illustrative example, a window of size $2$ ($N =
 1$), making Eq.(\ref{key}) read
 \begin{equation}
 \sum_{i,j=1}^{l} p(\omega_{i},\omega_{j})^{q}=
 \frac{1}{M}\sum_{j=1}^{M}\tilde{p}(x_{j},x_{j+1})^{q-1}.
 \label{windowofsize2}
 \end{equation}
 We note that
 \begin{equation}
 \tilde{p}(x_{j},x_{j+1}) = \rho(x_{j},x_{j+1}) (2/l)^{2},
 \label{simplecase}
 \end{equation}
 where $\rho(x_{j},x_{j+1})$ denotes the probability density at
 $(x_{j},x_{j+1})$.

 We show that the relation between the probability density
 referring to the window
 of size $N=2$ and that concerning the window of size $N=1$ is given by
 \begin{equation}
 \rho(x_{j},x_{j+1}) (2/l)^{2} = \rho(x_{j}) (2/l) exp(-\lambda_{j}).
 \label{densitydensity}
 \end{equation}
 To explain the dynamical origin of
 Eq.(\ref{densitydensity}) and to define the parameter $\lambda_{j}$ as
 well,
 it is convenient to remind the reader that, as earlier pointed out,
 we are considering a one-dimensional map,
 defined by
 \begin{equation}
 x_{j+1}=\Phi(x_{j}),
 \label{map}
 \end{equation}
 where, for the time being, the form of the function $\Phi(x)$ is left
 unspecified. The parameter $\lambda_{j}$ is the Lyapunov coefficient
 corresponding to the time step $j$, and it is defined as follows
 \begin{equation}
 \lambda_{j}\equiv ln(\frac{\Delta x_{j+1}}{\Delta x_{j}}),
 \label{lyapunov}
 \end{equation}
 or, equivalently, as
 \begin{equation}
 \lambda_{j}\equiv ln[\Phi^{\prime}(x_{j})].
 \label{derivative}
 \end{equation}
 We note that, in general, in the case of the intermittent map that we
 are considering, the Lyapunov coefficient does not lose dependence
 on the initial condition of the trajectory.The
 initial point of a trajectory is, let us say, $x_{0}$ and
 the Lyapunov coefficient is evaluated by studying, in addition to
 the trajectory starting at $x_{0}$, also an auxiliary trajectory
 which begins at $x_{0} + \Delta x_{0}$. The symbol $\Delta x_{j}$
 denotes the distance between the two trajectories at the time $j$. We shall
 come back to the important aspect of the dependence on the initial
 condition later on in this letter. From the definition itself of
 Eq.(\ref{lyapunov}) we have:
 \begin{equation}
 \Delta x_{j+1} = \Delta x_{j}exp( \lambda_{j}t).
 \label{dilatation}
 \end{equation}
 The departure of a trajectory from another one, initially belonging to
 the same interval of size $1/l$, implies that the total number of trajectories
 contained in the square of area $l^{-2}$ is decreased by the exponential
 $exp(-\lambda_{j})$, thereby  resulting in Eq.(\ref{densitydensity}) as
 an effect of moving from a
 window of size $1$ to a window of size $2$.

 It is evident that the counterpart of
 Eq.( \ref{simplecase}) in the case of a window of size $1$
 is $\tilde{p}(x_{j}) =  \rho(x_{j}) (2/l)$.
Thus Eq.(\ref{densitydensity}) becomes
 \begin{equation}
 \tilde{p}(x_{j},x_{j+1}) =  \tilde{p}(x_{j})exp(-\lambda_{j}).
 \label{example}
 \end{equation}
 The effect of moving from the window of size $1$ to a generic window of
size $N$ is
 expressed by the more general equation
 \begin{equation}
 \tilde{p}(x_{j},\ldots, x_{j+N}) =
  \tilde{p}(x_{j})exp(- \sum_{n=0}^{N-1}\lambda_{j + n}).
  \label{generalequation}
  \end{equation}

  Before illustrating the physical  consequences of the general
  prediction of Eq.(\ref{generalequation}) it is convenient to make
  some more comments about the Lyapunov coefficient time evolution.The
  conclusion expressed by Eq.(\ref{generalequation}), supplemented
  by Eq.(\ref{derivative}), leads to define the following time dependent
  Lyapunov coefficient:
  \begin{equation}
  \Lambda(N,x) \equiv  \sum_{n=0}^{N-1}ln[\Phi^{\prime}(x_{n})]
  = \sum_{n=0}^{N-1}ln(\frac{\Delta x_{n+1}}{\Delta x_{n}})
  = ln(\prod_{n=0}^{N-1}\frac{\Delta x_{n+1}}{\Delta x_{n}})
  =ln(\frac{\Delta x_{N}}{\Delta x_{0}}).
  \label{equationchain}
  \end{equation}
  It is convenient to express the property of an auxiliary trajectory
  of departing from the trajectory under study through the following
  function $\delta(t)$
  \begin{equation}
  \delta(t) \equiv
 lim_{\Delta x_{0} \rightarrow 0} \frac{\Delta
  x_{t}}{\Delta x_{0}}.
  \label{deltadefinition}
  \end{equation}
   Note that we are now considering windows of a so large size $N$ as
  to identify $N$ with the continuous variable $t$ and to neglect $1$
  compared to $N$. Tsallis et al.\cite{TPZ97} have recently shown
  that the extension of the KS method, resting on the Tsallis entropy
  of Eq.(\ref{entropy}), yields
  \begin{equation}
  \delta(t) = [1 + (1-Q)\kappa_{Q}(x)t]^{ \frac{1}{1-Q}}.
  \label{heuristic}
  \end{equation}
  This very important result was found by these authors studying the
  time evolution of the trajectories in the phase space corresponding
  to a window of size $1$, which corresponds to the one-dimensional phase
  space under study in this paper. Then these authors made the
  heuristic assumption that the increase of the number of cells takes
  place according to the same prediction as that leading to the
  increasing departure of a trajectory of interest from an auxiliary
  trajectory, namely, a trajectory with an initial condition
  very
  close to that of the trajectory of interest. This assumption is the same
as that
  adopted in this paper to derive Eq.(\ref{densitydensity}).
 They also made the heuristic assumption, not done here, that at each
  time step the occupied cells have the same probability.
  The validity of the prediction of Eq.(\ref{heuristic}) has been checked
in two distinct cases.
  The former case is that of the logistic map at the onset of chaos, where
the adoption of multifractal arguments\cite{lyra} yields $Q <1$.
  It has to be pointed out, however, that Lyra and Tsallis\cite{lyra}
   do not explain why the numerical calculation of the function $\delta(t)$ of
  Eq.(\ref{deltadefinition}) results in wild fluctuations and
   Eq.(\ref{heuristic}) is found to be accurate
  only for the amplitude increase of these oscillations.
  The latter case refers to the maps used to derive  L\'{e}vy
  diffusion processes\cite{ALLEGRO}. In this case the mere adoption
  of analytical calculation yields\cite{ANNA} $Q>1$.

  The important research work of Lyra and Tsallis\cite{lyra} and Tsallis
  and co-workers\cite{TPZ97} did not afford direct indications of how to
  carry out in practice the calculation of the KST entropy, and,
  especially,
  of how to go through a kind of statistical averaging.
 We are now in the right position to provide these directions
 so as to discuss to what an extent the entropic indexes $Q$
 predicted by the earlier papers of Refs.\cite{lyra},\cite{TPZ97}
 and\cite{ANNA} really result in a linear time increase of the KST entropy.
 In fact from the joint use
  of Eq.(\ref{generalequation}), Eq.(\ref{equationchain})  and
  Eq.(\ref{deltadefinition}) we get:
  \begin{equation}
  H_{q}(t) = [1 - k(q) \int dx p(x)^{{q}}\delta(t,x)^{1-q}]/(q-1),
  \label{fundamentalresult}
  \end{equation}
  where $k(q) \equiv  (2/l)^{q-1}$ stems from the replacement of the
  discrete sum of Eq.(\ref{key}) with the continuous integral. Note that
  this makes the KST entropy dependent on the size of the cells as the so
  called $\epsilon-entropy$ (see, for example,
  Ref.\cite{wangalone}).Note also that to derive
  Eq.(\ref{fundamentalresult}) we adopted the arguments
  of Ref.\cite{chinesebook}, and thus
  the obvious generalization
  of Eq.(\ref{key}):

   \begin{equation}
 \sum_{i} p_{i}^{q} f(x_{i})=
 \frac{1}{M}\sum_{j=1}^{M}\tilde{p}_{j}^{q-1}f(x_{j}),
 \label{keycopy}
 \end{equation}
 where $f(x_{i})$ is a generic function,

  It is evident that the prediction of Refs.\cite{lyra},\cite{TPZ97},
  yielding $Q<1$, results in a linear time increase of the KST entropy
  only if the fluctuations can be ignored. Actually these fluctuations are
  multifractal in nature and are
  a manifestation of the effect observed more than ten years ago by
  Anania and Politi \cite{politi}. These authors showed
  that the Feigenbaum attractor, discussed in terms
  of an algebraic index $\beta$, results in a fluctuating
  spectrum $h(\beta)$.
  They also noticed that the behavior of a finite distance is described
  by algebraic exponents over a a limited range.
  All these observations might be related
  to the possibility that the entropic index $Q$
  has to be considered as time dependent also in the
  case $Q < 1$. We are confident that
  the result of Eq.(\ref{fundamentalresult}) affords a way of
  discussing the thermodynamic aspects of these interesting
  phenomena.

  Let us now apply the fundamental result of Eq.(\ref{fundamentalresult})
  to the case of intermittent motion.  This is the Manneville
map\cite{manneville}, whose
  explicit expression is:
  \begin{equation}
  x_{n+1} = \Phi(x_{n}) = x_{n} + x_{n}^{z} (mod.1) (1\leq z).
  \label{manneville}
  \end{equation}
  This map has been more recently used by Gaspard and
  Wang\cite{Gaspardwang} to discuss the algorithmic complexity of
  sporadic randomness. Using the main result of this paper we
  are now in a position to prove the aging effect of the
  non-extensive thermodynamics of Tsallis. We limit our predictions
  to three distinct time  regions. It has to be stressed that a first
  time scale is given by the time step $\Delta t = 1$. The inverse
  power law nature
  of the waiting time distribution
  $\psi(t)$ is perceived at times much larger than $\Delta t =1$.
  The short-time regions is given by times comparable to this
  microscopic time scale, which is expected to be dominated
  by a condition of total chaos, and, consequently to be associated with
  $Q = 1$. In fact the short-time region will be dominated by the
  trajectories with initial conditions in either the chaotic region or
  in the laminar region, but conveniently close to the border with the
  chaotic region. Another important time is given by $T$, which
  is the mean waiting time in the laminar region, finite
  for $z<2$. The intermediate time region refers to times
  $\Delta t  << t<<T$  and the large time regions refers to $t>>T$.

  In the intermediate time region adopting an approach similar to that used
  in Ref.\cite{ANNA}, we find that the heuristic prediction
  of Eq.(\ref{heuristic}) is exact. Furthermore we find :
  \begin{equation}
  Q = 1 + (z-1)/z
  \label{magicq}
  \end{equation}
  and
  \begin{equation}
  \kappa_{Q}(x) = \frac{x^{\frac{Q-1}{2-Q}}}{2-Q}.
  \label{xdependentlyapunov}
  \end{equation}
  In conclusion we obtain that when the entropic index $q$ is given the
  magic value established by Eq.(\ref{magicq}), the time evolution of
  $H_{q}(t)$ becomes linear. Thus we recover the prediction of
  Ref.\cite{ANNA}, implying that the dynamical processes of
  L\'{e}vy diffusion, in an intermediate time region, are
  associated to non-extensive thermodynamics. This is not
  a permanent condition: We know from the work of Gaspard and
  Wang\cite{Gaspardwang} that in the long-time limit ($N \rightarrow
  \infty$)
  $\Lambda(N,x)/N$ becomes constant and loses the dependence on the
  initial condition. Consequently, on the basis itself of
  the theory illustrated in this paper, it is straigtforward to make
  the prediction that in the long-time limit $Q=1$.

  It is also evident that the theoretical predictions of this paper
  refer to a case where the size $1/l$ of the cells is made
  arbitrarily small. This explains why in the intermediate time region
  $Q = 1 + \frac{z-1}{z} > 1$,  in an apparent conflict with the result of the
  numerical analysis of Ref.\cite{buiatti} yielding $ Q =
  [\frac{z}{(z-1)}- 2]^{\alpha}$ with $\alpha \approx 0.15$, namely,
  $Q < 1$. This is so because the authors
  of Ref.\cite{buiatti} studied the KST entropy
  of a dynamical system with algoritmic complexity
  equivalent to that of the map of Ref.\cite{geisel}, and so to that
  of the the Manneville map, using only two cells. The corresponding
  symbolic sequence
  is characterized by extended strings with the same symbol, either
  $1$ or $-1$,
  and the entropy increase is generated by the
  random length of these sequences. The strings with
  the same symbol are indistinguishable from phases
  of regular motion, implying $Q =0$, while the sporadic randomness
  would yield $Q=1$. Therefore, it is reasonable that the numerical
  analysis, as a balance between these two processes, yields $ Q <1$.
 The adoption of arbitrarily small cells, on the
  contrary, makes it possible to relate the extended laminar regions
  to $Q >1$, in accordance with the prediction on superdiffusion
  of\cite{buiatti}.

  We notice that the nature itself of the connection between dynamics
  and thermodynamics established by the work of
  Refs.\cite{lyra} and \cite{TPZ97} implies the aging of
  the non-extensive thermodynamics of superdiffusion. In fact, in
  this case $Q>1$ and Eq.(\ref{heuristic}) yields a function
  $\delta(t)$ faster than the exponential, in the sense that
  this function diverges at a finite time. Actually, this divergence
  does not have a physical significance: The function $\delta(t)$ is
  forced to depart from the prediction of Eq.(\ref{heuristic}) by the
 exit from the laminar region. In the long-time scale the sequel of
 many exits from the laminar region and many random injections into to
 it makes it possible to adopt Eq.(\ref{heuristic}), provided
 that the entropic index $Q$ is assumed to slowly regress to the
 ordinary value $Q = 1$.We think
  that this conclusion agrees with the observation made by
Wang\cite{wangalone} that
  the $\epsilon$-entropy increases linearly in time for a L\'{e}vy
  process. In fact, from the entropic point of view the Manneville map is
  equivalent to the map used in \cite{ALLEGRO} for the dynamical
  derivation of the L\'{e}vy processes, and Ref.\cite{ALLEGRO}, in turn,
  shows that the L\'{e}vy diffusion regime is reached as a consequence of
  the repeated action of randomness, established
  by the chaotic part of the map. The
  ultimate effect of this randomness is that of producing the Markov
  property, a condition necessary for the
  realization of the L\'{e}vy diffusion, which is in fact Markov.
 In principle, the case $Q<1$ might be compatible with an eternal form
 of non-extensive thermodynamics. If this is so, or not for reasons
 related to the effects discovered by Anania and
 Politi\cite{politi},can be
 assessed with further research work based on
 the adoption of the important result of
 Eq.(\ref{fundamentalresult}).

  We are now in a position to answer the important issue raised by
  Zaslavsky.
  It is easy to relate all these results to the Hamiltonian dynamics
  of interest for Zaslavsky\cite{zaslavsky}. In fact, as pointed out
  by Zaslavsky himself, the main statistical properties of the
  dynamical processes of interest are determined by the waiting time
  distribution $\psi(t)$ of Eq.(\ref{inversepower}) and the time evolution of
  the Lyapunov coefficients is strictly dependent on the power law of
  $\psi(t)$, as it can be realized by comparing
  Eq.(\ref{inversepower}) to Eq.(\ref{heuristic}) in the light
  of the conjecture made in Ref. \cite{ANNA} that
   $\psi(t)= k \delta(-<\kappa_{0}(x)>t)$. We also note that
  it is straigthforward to prove that the KST entropy of the billiards
discussed by
  Zaslavsky can be evaluated using the prescription of Eq.
  (\ref{fundamentalresult}), provided that the probability
  distribution $p(x)$ is meant to refer to the corresponding
  phase space. Zaslavsky shows that the short-time evolution of
  $\psi(t)$ is Poisson-like.
  We note also that the conjecture
 $\psi(t)= k \delta(-<\kappa_{0}(x)>t)$
  fits the prescription of Eq. (\ref{bernouilli}) in the case
  where $Q = 1$. Thus, if we limit our observation to the short-time
  dynamics we reach the conclusion that the entropic index
  fits the extensive requirement $Q = 1$. At later times, however,
  when the inverse power law nature of the function
  $\psi(t)$ shows up, we expect  that $Q$ might come close to the
  non-extensive value $Q = 1 + 1/(2 + \beta)$ resulting from the theoretical
  analysis of Ref.\cite{ANNA}. Furthermore, on the basis of the  results
  of Ref.\cite{Gaspardwang}, we expect again that in the long-time
  limit the extensive value $Q=1$ is recovered.
 The billiards studied by Zaslavsky\cite{zaslavsky} are characterized
 by the joint action of a chaotic sea and of the fractal dynamics at
 the border between chaotic sea and stability islands. We
  believe that the memory erasing process that according to Zaslavsky is
  necessary to suppress the effects of Maxwell's demon is produced
  by the action itself of the chaotic sea. The extended time regime
  prior to this final condition, however, is already thermodynamical,
  and this paper answers the question raised by Zaslasky about the
  thermodynamic nature of a dynamical system whose statistical
  properties seem to be a manifestation of Maxwell's demon. The main
  conclusion of this letter is that the Maxwell demon is
  compatible with
  thermodynamics provided that the non-extensive
  perspective of Tsallis is adopted.

\end{document}